# AN EMPIRICAL STUDY OF OPEN SOURCE SOFTWARE USABILITY:
## THE INDUSTRIAL PERSPECTIVE

*Arif Raza, University of Western Ontario, Canada*
*Luiz Fernando Capretz, University of Western Ontario, Canada*
*Faheem Ahmed, United Arab Emirates University, UAE*

## ABSTRACT

Recent years have seen a sharp increase in the use of open source projects by common novice users; Open Source Software (OSS) is thus no longer a reserved arena for software developers and computer gurus. Although user-centered designs are gaining popularity in OSS, usability is still not considered as one of the prime objectives in many design scenarios. In this paper, we analyze industry users' perception of usability factors, including understandability, learnability, operability and attractiveness, on OSS usability. The research model of this empirical study establishes the relationship between the key usability factors and OSS usability from industrial perspective. In order to conduct the study, a data set of 105 industry users is included. The results of the empirical investigation indicate the significance of the key factors for OSS usability.

## INTRODUCTION

In the ISO 9241-11 (1998) standard, usability is defined as *"the extent to which a product can be used by specified users to achieve specified goals with effectiveness, efficiency and satisfaction in a specified context of use."* However, The International Organization for Standardization and The International Electro technical Commission ISO/IEC 9126-1 (2001) categorizes software quality attributes into six categories: namely functionality, reliability, usability, efficiency, maintainability and portability. In the standard, usability is defined as *"the capability of the software product to be understood, learned, used and attractive to the user, when used under specified conditions."* Here, usability is further subdivided into understandability, learnability, operability and attractiveness.

While studying GNOME project, Koch and Schneider (2002) observe that in general, the number of people involved in OSS development are more than in traditional organizations, *"but the data show the existence of a relatively small 'inner circle' of programmers responsible for most of the output."* OSS users, however, come from every corner of the world having all sort of cultural, technical and non-technical backgrounds, requirements and expectations. They have free access as well as the ability to modify the source code (Aberdour, 2007). OSS is no longer reserved for computer developers alone, since a number of non-technical and novice computer users are growing at a fast pace, underscoring the need to understand and address their requirements and expectations (Iivari, 2009a). Although Laplante et al. (2007) believe that OSS has more potential to achieve higher software quality as compared to closed proprietary software; they observe the reluctance shown by many organizations in using OSS primarily due to *"an inherent distrust of OSS quality."* Nichols and Twidale (2006) state, *"it is unfair to compare imperfect but public OSS processes with imagined but concealed commercial processes."* They believe that due to the OSS environment, the software development process has become accessible that has been kept concealed in proprietary software. Referring much of the commercial software that failed to address usability issues properly, the authors do not consider usability a resolved issue in closed software projects either. They believe that research in the domain of OSS usability would be beneficial to both OSS as well as closed proprietary software products. Hedberg et al. (2007) observe that with the rapid increase in the non technical users of OSS, expectations related to higher software quality will grow as well. According to them, unlike the typical OSS approach, users will not be the co-developers who are competent enough to locate and fix the bugs; thus the quality assurance would need to be done before the software is delivered. They stress the need of having empirical research dealing with usability and quality assurance in OSS. de Groot et al. (2006) maintain that *"many OSS projects, such as KDE, have established processes for the maintenance of software quality. However, these can only be of limited use when the actual quality of the product is still unknown."* While carrying out a study on the evolution metrics of OSS, Wang et al. (2007) propose a new set of metrics. Furthermore, their case study on Ubuntu – a popular Linux distribution, confirms the essential role of open source community and its members in OSS evolution.

Winter et al. (2007) consider the improvement of *"the usage of a system"* to support user activities as the main aim of usability engineering. Bodker et al. (2007) highlight that OSS developers need to have a full understanding, motivation and determination to address users' demands to avoid ending up with products that lack user friendliness, which could be a serious threat to its popularity and adoption. Ahmed (2008) refers to questionnaires that have long been used to gather users' assessment regarding subjective matters such as interfaces. However he realizes the need of more resources for usability testing as its success relies upon the test quality and coverage. Zaharias and Poylymenakou (2009) also consider usability questionnaires as a fast, cost effective way to collect users' feedback that can also be used to confirm target users.

We have already conducted three studies to empirically investigate the significance of certain key factors on OSS usability from OSS developers, users and contributors (that include users, developers, testers, systems analysts) points of view. This research work is the last of the series in which we analyze the industry users' perception regarding impact of the sub-factors of usability (understandability, learnability, operability and attractiveness) upon OSS usability. This study contributes to understanding the effects of the stated key factors which play a vital role in OSS usability.

In Section-2, we present the literature review regarding software usability issues in the open source software industry, in general and related to the key factors considered in this study, in particular. In Section-3 the research model and the hypotheses of this study are presented. The research methodology, data collection process, and the experimental setup are explained in the first part of Section-4, reliability and validity analysis of the measuring instrument in the second; and data analysis procedures in its third part. In Section-5, hypotheses testing and the analysis of the results is presented. It is followed by the discussion in Section-6 that also includes the limitations of the study. Finally Section-7 concludes the paper.

## LITERATURE REVIEW

**Usability Issues – In General**

Golden (2009) observes that *"systems continue to be built and released with glaring usability flaws that are costly and difficult to fix after the system has been designed and/or built."* He stresses that addressing usability issues at a software architecture design level makes it cost effective for software developers.

Cox (2005) identifies the fact that although issues related to human factors and usability are considered, they are too late in the software life cycles to have any useful impact. Juristo (2009) also believes in considering usability earlier in the life cycle. She has also come up with an approach to incorporate usability features as functional requirements.

Fitzpatrick and Higgins (1998) have considered usable software in compliance with the latest legislation as the one that is demanded by end users. They stress the need of having clear attribute listing of usable software along with the applicable measuring procedures.

Chrusch (2000) refers to and negates the seven myths of usability such as software development cost and time both increase due to usability, user-interface is just about addition of graphics to make it attractive, usability is about the interface design alone or it is only about common sense, for good user interface design developer's familiarity with the standard guidelines is the only requirement, there is no need to do usability testing as long as developers have been working with the users long enough, and the final myth that usability issues will be addressed during help/documentation and training.

Te'eni (2007) believes that how useful and easy to use a system is, has a major role in determining user's intentions to make use of it.

Lewis (2006) stresses the designing systems for a wide range of users by stating that *"while public attitudes are improving, and integration into society of people with cognitive disabilities is increasing, there is still widespread ignorance about them and how technology can be of value to them."*

Seffah (2003) believes that software developers' knowledge regarding user interface design need to be enhanced to the level that they could be able to integrate such usability techniques in their design and development processes. In another work, Seffah and Metzker (2004) identify that due to the inconsistency in defining usability by standardization organizations and the software development industry, usability has different interpretations by different researchers. They stress the increase of communication between the software developers and the usability experts.

Advocating the concept of *"Universal Usability"*, Shneiderman (2000) observes that to accommodate a wide variety of users, researchers and designers have to come up with such innovative designs that could be beneficial to all sections of users. He stresses the development, testing and refinement of such universal software to address usability issues related to diverse set of users. According to him, *"reaching a broad audience is more than a democratic ideal; it makes good business sense."*

**Usability in Open Source Software**

Considering usability as a research area in OSS that needs to be examined, Hedberg et al. (2007) state *"user feedback should be sought early, and the design solution should be iterated based on the user feedback."* They see a great potential for usability experts to contribute towards OSS development.

According to Nichols and Twidale (2006), *"research in open source usability has the potential to be valuable to all kinds of software development, not just OSS."* They emphasize on finding ways to ease usability bug reporting as well as involving usability experts during software analysis and design phases.

Nichols et al. (2001) identify the inability of many OSS users to do debugging of source code and their need of support even in bug reporting. They maintain that *"as work on open-source projects is voluntary then developers work on the topics that interest them and this may well not include features for novice users."*

Iivari and Iivari (2006) realize that in most of the cases, neither the prospective users of a software product are known nor can they be involved individually, particularly if the users are geographically and organizationally distributed; as a result *"user focus can be limited to focus on typical, average or fictive user."*

In their empirical study, Andreasen et al. (2006) found that although OSS developers realize the importance of end users, usability related issues do not get top position in their priority list. They identify that *"currently, most developers have a very limited understanding of usability. Moreover, there is a lack of resources and evaluation methods fitting into the OSS paradigm."*

Çetin and Göktürk (2008) believe that high usability of an OSS project can only be achieved through its measurement and analysis. They propose a measurement framework to assess OSS projects, which is required for their self evaluation.

Referring to the structured defect handling processes, significant use of configuration and bug tracking tools, Otte et al.(2008) highlight high rate of user participation, user testing and peer reviews in OSS projects.

Lee et al. (2009) recommend in their empirical study that *"usefulness, ease of use, and reliability"* are some of the major factors that OSS practitioners shall pay attention to, for improving OSS quality.

**Literature Review of Key Factors**

Referring to the difficulties in usability testing, Lindgaard (2006) states that *"it is impossible to know whether all usability problems have been identified in a particular test or type of evaluation unless testing is repeated until it reaches an asymptote, a point at which no new problems emerge in a test."* Iivari (2009b) empirically studies user participation in an OSS project and acknowledges *"informative, consultative and participative roles for users."* Viorres et al. (2007) believe in the end-users involvement during software design and development. They recognize the need of applying human-computer interaction (HCI) principles in the design processes of OSS to make use of their full potential.

According to Seffah et al. (2006), failure of most interactive systems is mainly due to the unusable user interfaces. Referring to the difficulties in measuring software usability, they highlight the need to know the characteristics of users, their intended tasks, and identify that the lacking of either of the factors would end up in unrealistic results.

In ISO/IEC 9126-1 (2001), understandability is defined as *"the capability of the software product to enable the user to understand whether the software is suitable, and how it can be used for particular tasks and conditions of use."* According to Seffah and Metzker (2004), software developers and usability experts can both be benefited, if they understand culture and practices of HCI and software engineering (SE), and learn techniques to improve communication between the two disciplines. Mørch et al. (2004) realize that understandability of end users can be increased, if developers could understand the semantics of integrating different user interface components. Highlighting the diversity, different intelligence levels, and approaches of end users; Shneiderman (2000) states that some need less and some need more time to understand and acquire knowledge about new tools and user interface. Hedberg et al. (2007) refer to multiple meanings of user centered design (UCD) methodology; they argue that all of them *"emphasize the importance of understanding the user, his/her tasks or work practices and the context of use."*

Learnability is defined in ISO/IEC 9126-1 (2001) as *"the capability of the software product to enable the user to learn its application."* Seffah et al. (2006) identify the need of more comprehensive guidelines to *"account for the degree of influence of individual quality factors, such as the role of learnability versus understandability in usability problems."* Mishra and Hershey (2004) stress that the understanding of requirements and knowledge background of users, can develop better learning tools. Yunwen and Kishida (2003) consider learning as one of the main motivational forces, which results in the participation of both users as well as software developers, in OSS culture. They believe that new members and users are attracted to OSS because of its high quality; as one of their own problems could be solved by the system whereas developers are attracted to OSS due to its learning opportunities.

Operability is defined in ISO/IEC 9126-1 (2001) as *"the capability of the software product to enable the user to operate and control it."* Henderson (2005) emphasizes that developers should produce software having usable interface, which could meet user needs, and provide them the value they expect. Iivari and Iivari (2006) state that ideally an individual's needs should be supported by a system; they, however, realize that in real world each and every user cannot be accessed while designing, plus users should be prepared to make some compromises to have a uniform and a compatible system. The authors state that *"in certain situations the prospective users can all participate directly in the process, but in many cases only selected user representatives are involved."* Crowston et al. (2003), while discussing the success of open source software projects, identify the contrasting features of proprietary software and OSS. They agree that in either case, system's success is measured through user satisfaction, most OSS projects are globally distributed with unknown population of users, which makes it hard to have true sample of users.

ISO/IEC 9126-1 (2001) defines attractiveness as *"the capability of the software product to be attractive to the user."* Chrusch (2000) believes that proper application of usability techniques results in a good user interface. He observes that *"many people misinterpret the visual design of an interface as the interface itself, but doing so ignores the entire interaction sequence needed to complete a task."* Juristo (2009) identifies a flaw in the approach of development team, when they think that a system can be made usable by incurring right font, color, and nice set of controls. Markov (2003) states that usability is not about making a user interface attractive, rather it is about *"total user experience."*

## RESEARCH MODEL AND THE HYPOTHESES

In this study we present a research model to analyze and empirically investigate the relationship between the key usability factors and the open source software usability. The theoretical model to be empirically tested in this study is shown in Fig. 1.

We will examine the relationship of four independent variables and the OSS usability, which is the dependent variable in this model. Our aim is to investigate the answer to the following research question:

**Research Question**: *How do usability components (understandability, learnability, operability and attractiveness) affect usability from the industry users' perspective?*

There are four independent and one dependent variable in this research model. The four independent variables are called "Usability Factors" in the rest of the paper. They include Understandability, Learnability, Operability and Attractiveness. The dependent variable of this study is the OSS usability. The multiple linear regression equation of the model is as follows:

$$\text{OSS Usability} = f_0 + f_1 v_1 + f_2 v_2 + f_3 v_3 + f_4 v_4 \quad (1)$$

where $f_0, f_1, f_2, f_3$ and $f_4$ are the coefficients and $v_1, v_2, v_3$ and $v_4$ are the four independent variables. In order to empirically investigate the research question, we hypothesize the following:

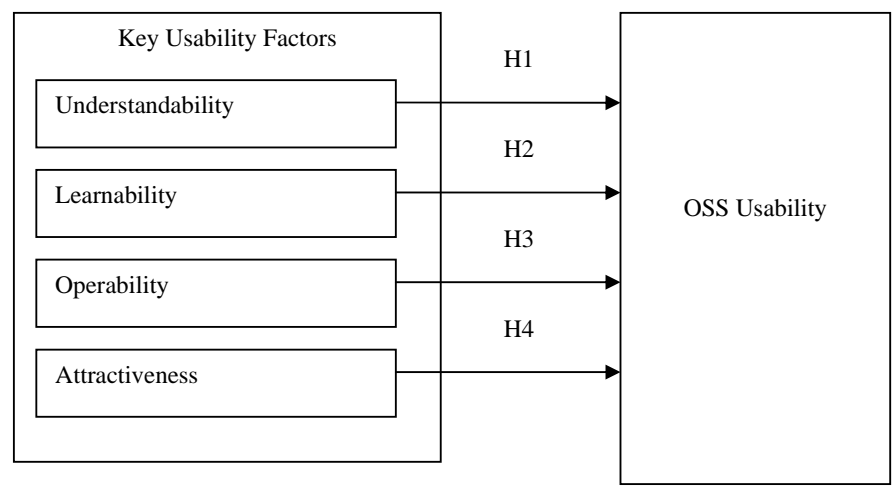

Fig.1 Research Model

**H1:** **Understandability** is positively related with OSS usability.

**H2:** **Learnability** has a positive impact on OSS usability.

**H3:**   **Operability** plays a positive role towards usability in OSS.

**H4:**   **Attractiveness** is positively related with OSS usability.

## RESEARCH METHODOLOGY

The research conducted and presented in this paper includes the empirical results of a survey. In this study, the target population includes multinational companies whose employees are OSS users. Thirty companies consented to participate in this study, with the assurance of confidentiality for both the organization and the individuals. The participating organizations are involved in a wide range of operations, such as pharmaceuticals, telecommunications, automobile manufacturing, information technology and consumer electronics. Specifically, these organizations include North American and European multinational companies, and they vary in size from small to large scale. We requested that the companies in the study distribute the questionnaires within their various departments, so that we have several responses from within the same organization. In particular, we required that the respondents possessed the minimum educational qualification of an undergraduate degree.

The survey was implemented by using the survey tool *"kwiksurveys"*. It was started in the last week of March 2010, and it was closed after three weeks, with 105 responses. We assured the participants that our survey neither required their identity nor would be recorded. However, to support our data analysis of the respondents' experience, we asked them, *"Do you agree that applying one of the concepts/techniques expressed by the above key factors, usability will, in your opinion, improve the product you are working on?"* Out of 105 total responses, 81% agreed that in their experience, the application of our key factors will improve the usability of their application; of the remaining participants, 16% were neutral and 3% disagreed with this statement, as reflected in Figure 2.

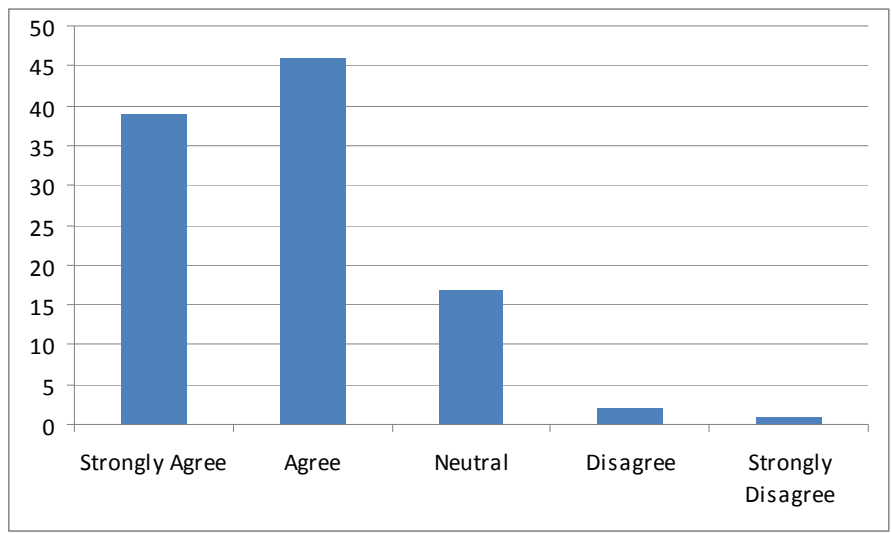

Fig. 2 –Application of Usability Factors in Respondents' products.

**Data Collection and the Measuring Instrument**

In this study, the questionnaires presented in Appendix A were used to learn, up to what extent these usability factors were important towards OSS usability, for the respondents of the survey. The questionnaires required the respondents to indicate the extent of their agreement, or disagreement with statements using a five-point Likert scale. We used sixteen separate items to measure the independent variables, and four items to measure respondents' points of view regarding OSS usability. We reviewed previous researches on the subject of OSS usability, so that a comprehensive list of measuring factors could be constructed. To measure the extent to which each of these usability factors have been practiced in their projects, we made use of five-point Likert scale. The Likert scale ranged from "Strongly Agree" (1) to "Strongly Dis-agree" (5), for all items associated with each variable. The items for all the four usability factors are labeled sequentially in Appendix A and are numbered 1 through 16. We measured the dependent variable, i.e. OSS Usability on the multi-item, five-point Likert scale too. The items were specifically designed, for collecting measures for this variable, and are labeled sequentially from 1 through 4 in Appendix A.

**Reliability and Validity Analysis of Measuring Instrument**

The two integral features of any empirical study are reliability, which refers to the consistency of the measurement; and the validity, which is the strength of the inference between the true value and the value of a measurement. For this empirical investigation, we used the most commonly used approaches in empirical studies, to conduct reliability and validity analysis of the measuring instruments. The reliability of the multiple-item measurement scales of the four usability factors was evaluated by using internal-consistency analysis, which was performed using coefficient alpha (Cronbach, 1951). In our analysis, the coefficient alpha ranged from 0.70 to 0.73 as shown in Table 1. Nunnally and Bernste (1994) find that a reliability coefficient of 0.70 or higher for a measuring instrument is satisfactory. van de Ven and Ferry (1980) state that a reliability coefficient of 0.55 or higher is satisfactory, and Osterhof (2001) suggests that 0.60 or higher is satisfactory. Therefore, we concluded that the variable items developed for this empirical investigation were reliable.

Table 1: Coefficient alpha and Principal Component Analysis (PCA) of variables

| Usability Factors | Item no. | Coefficient α | PCA Eigen value |
|---|---|---|---|
| Understandability | 1 - 4 | 0.73 | 1.86 |
| Learnability | 5 - 8 | 0.72 | 1.48 |
| Operability | 9 - 12 | 0.70 | 1.01 |
| Attractiveness | 13 - 16 | 0.71 | 1.06 |

Comrey and Lee's (1992) Principal Component Analysis (PCA) was performed for all the four key usability factors, and reported in Table 1. We used Eigen value (Kaiser, 1970) as a reference point, to observe the construct validity, using principal component analysis. In this study, we used Eigen value-one-criterion, also known as Kaiser Criterion (Kaiser, 1960; Stevens, 1986), which means any component having an Eigen value greater than one was retained. Eigen value analysis revealed that all the four variables completely formed a single factor. Therefore we concluded that the convergent validity, sufficient for the data.

**Data Analysis Procedure**

We analyzed the research model, and the significance of hypotheses H1-H4, through different statistical techniques in three phases. In phase-I we used normal distribution tests and parametric statistics, whereas in phase II we used non-parametric statistics. Due to the relatively small sample size, both parametric as well as non-parametric statistical approaches were used, to reduce the threats to external validity. As our measuring instrument had multiple items for all the four independent variables as well as the dependent variable (refer to Appendix A), their ratings by the respondents were summed up, to get a composite value for each of them. Tests were conducted for the hypotheses H1-H4, using parametric statistics by determining the Pearson correlation coefficient. For non-parametric statistics, tests were conducted for the hypotheses H1-H4, by determining the Spearman correlation coefficient. To deal with the limitations of the relatively small sample size and to increase the reliability of the results, the hypotheses H1-H4 of the research model were tested, using Partial Least Square (PLS) technique in Phase-III. According to Fornell and Bookstein (1982) and Joreskog and Wold (1982), the PLS technique is helpful in dealing with issues such as complexity, non-normal distribution, low theoretical information, and small sample size. The statistical calculations were performed using minitab- 15.

## HYPOTHESES TESTING AND RESULTS

**Phase-I**

To test the hypotheses H1-H4 of the research model (shown above in Fig. 1), parametric statistics were used in this phase by examining the Pearson correlation coefficient between individual independent variables (key usability factors) and the dependent variable (OSS usability). The results of the statistical calculations for the Pearson correlation coefficients are displayed in Table 2. The Pearson correlation coefficient between Understandability and OSS usability was found positive (0.42) at $P < 0.05$, and hence justified the hypothesis H1. The Pearson correlation coefficient of 0.42 was also observed at $P < 0.05$ between Learnability and OSS usability, and hence found significant at $P < 0.05$, as well. The hypothesis H3 was accepted based on the Pearson correlation coefficient (0.51) at $P < 0.05$, between Operability and OSS usability. The positive correlation coefficient of 0.40 at $P < 0.05$ was also observed between the OSS usability and Attractiveness, which meant that H4 was accepted too.

Hence, as observed and reported above all the hypotheses H1, H3 and H4 were found statistically significant and were accepted.

**Phase II**

Non-parametric statistical testing was conducted in this phase by examining Spearman correlation coefficients between individual independent variables (key usability factors) and the dependent variable (OSS usability). The results of the statistical calculations for the Spearman correlation coefficients are also displayed in Table 2. The Spearman correlation coefficient between Understandability and OSS usability was found positive (0.40) at $P < 0.05$, and hence justified the hypothesis H1. For hypothesis H2, the Spearman correlation coefficient of 0.41 was observed at $P < 0.05$, hence significant relationship was found between Learnability and OSS usability in this test. The hypothesis H3 was accepted, based on the Spearman correlation coefficient (0.51) at $P < 0.05$, between Operability and OSS usability. The positive Spearman correlation coefficient of 0.37 at $P < 0.05$ was also observed between the OSS usability and Attractiveness which meant that H4 was accepted too.

Hence, as observed and presented above all the hypotheses H1, H2, H3 and H4 were found statistically significant and were accepted in the non-parametric analysis as well.

Table 2: Hypotheses testing using parametric and non-parametric correlation coefficients

| Hypothesis | Usability Factor | Pearson Correlation coefficient | Spearman Correlation coefficient |
|---|---|---|---|
| H1 | Understandability | 0.42* | 0.40* |
| H2 | Learnability | 0.42* | 0.41* |
| H3 | Operability | 0.51* | 0.51* |
| H4 | Attractiveness | 0.40* | 0.37* |

* Significant at $P < 0.05$.

**Phase III**

In order to do the cross-validation of the results obtained in Phase I and Phase II, Partial Least Square (PLS) technique was used in this phase of hypotheses testing. The direction and significance of hypotheses H1–H4 were examined. In PLS, the dependent variable of our research model, OSS usability was placed as the response variable and the independent key usability factors as the predicate. The test results containing observed values of path coefficient, $R^2$ and F-ratio have been shown in Table 3. Understandability was observed to be significant at $P < 0.05$, with path coefficient 0.63, $R^2$: 24.9% and F-ratio as 27.54. Learnability had path coefficient of 0.85 with $R^2$: 65.3% and F-ratio of 156.44 and found significant at $P < 0.05$ as well. Operability was observed to have the same direction as proposed in the hypothesis H3, with path coefficient: 0.80, $R^2$: 55.6% and F-ratio: 103.9 at $P < 0.05$. And finally Attractiveness with the path coefficient: 1.08, $R^2$: 60.7% and F-ratio: 128.35 at $P < 0.05$, was also found in accordance with the hypothesis H4.

Table 3: Hypotheses testing using Partial Least Square (PLS) regression

| Hypothesis | Usability Factor | Path Coefficient | $R^2$ | F- Ratio |
|---|---|---|---|---|
| H1 | Understandability | 0.63 | 0.249 | 27.54* |
| H2 | Learnability | 0.85 | 0.653 | 156.44* |
| H3 | Operability | 0.80 | 0.556 | 103.9* |
| H4 | Attractiveness | 1. 08 | 0.607 | 128.35* |

* Significant at $P < 0.05$.

### Testing of the Research Model

The multiple linear regression equation of our research model is depicted by Equation-1. The purpose of research model testing was to provide empirical evidence that our key factors play a significant role towards open source software usability. The testing process consists of conducting regression analysis, and reporting the values of the model coefficients, and their direction of association. We placed OSS usability as response variable and key factors as predicators. Table 4 displays the regression analysis results of the research model. The path coefficients of the four variables: understandability, learnability, operability and attractiveness were found positive, and their t-statistics were also observed statistically significant at $P < 0.05$. $R^2$ and adjusted $R^2$ of overall research model were observed as 0.40 and 0.373 with F-ratio of 14.97, significant at $P < 0.05$.

Table 4: Multiple Linear Regression Analysis of the research model.

| Model coefficient Name | Model coefficient | Coefficient value | t-value |
|---|---|---|---|
| Understandability | $f_1$ | 0.180 | 2.73* |
| Learnability | $f_2$ | 0.150 | 1.72* |
| Operability | $f_3$ | 0.154 | 2.12* |
| Attractiveness | $f_4$ | 0.238 | 2.85* |
| Constant | $f_0$ | 2.72 | 2.77* |

* Significant at $P < 0.05$.

## DISCUSSION – QUESTIONNAIRES AND RESPONSES

It is generally believed that testing procedures, in particular usability testing are conducted in different manners in closed proprietary software and in OSS projects. However many issues remain common in both. That is the reason some of the questions in our survey are specifically related to OSS and others are not, as we believe they are equally applicable to usability assessment in proprietary organizations as well as OSS projects. We have tried our level best to come up with open questions and avoid leading questions. As already mentioned, we have designed four items for each independent variable to collect measures on the extent to which the variable is practiced within each project.

In questions related to understandability, we have asked respondents' opinions about the relationship between understandability and functionality as well as about consistency and understandability. One of the statement related to Understandability (refer to Appendix A) *"Easy to understand software would encourage user's involvement"* we believe, is beyond the categorization of OSS or proprietary software users, however equally important for both. When we asked whether software functionality needs to be compromised in order to increase understandability, 72% of the respondents disagreed (either disagreed or strongly disagreed), 12% agreed (either agreed or strongly agreed) and 16% remained neutral. 79% agreed that consistency in software design would increase understandability and hence usability; 16% remained neutral and only 5% disagreed with the statement. When we asked respondents' opinion about the statement *"Easy to understand software would encourage user's involvement,"* 81% agreed, 13% remained neutral and 6% disagreed. And finally about the *"Inconsistency in software is due to a lack of understanding user's expectations"* (Chrusch, 2000), 37% agreed, 40% disagreed and the rest 23% chose to remain neutral. Thus, overall, as our statistical analysis indicates, our hypothesis *"H1: Understandability is positively related with OSS usability,"* is found significant and has been accepted in the analysis.

Regarding learnabilty, we have asked our respondents about the relationship between learnability, accessibility and usability. We also inquired about whether OSS developers compromise learnability for efficiency and whether learnability being considered impracticable in the OSS environment. Our final question about learnability was related to the realization of the fact that a system could be made learnable only if its developer understands the needs and limitations of its users. 83% agreed that learnability increases accessibility and hence usability, 14% remained neutral and 3% disagreed. We have also asked the respondents to opine whether learnability may be compromised by developers to produce efficient products in OSS environment; 34% agreed, 42% remained neutral and 24% disagreed with the statement. 76% disagreed on considering learnability as a cognitive issue that is not practicable in OSS, 18% were neutral and only 6% agreed. Regarding the statement that to make a system learnable, OSS developers must understand the limits of their target users 72% agreed, 14% remained neutral and 14% chose to disagree. On the basis of the statistical investigation, the hypothesis *"H2: Learnability has a positive impact on OSS usability"* has been accepted.

In order to keep our statements unbiased and open, we asked our respondents' opinion about whether they agree that more learnable software makes it more operable and usable. We also asked them to opine about gradual introduction of advanced features in software. Responding to our survey statements related to operability, 76% believed that more learnable software is more operable and hence usable, 13% remained neutral and the rest 11% disagreed. 63% agreed that introducing advanced features of software to users in an incremental way would give them more control in using the software; 20% remained neutral and 17% disagreed. There was a mixed response about the statement *"Operability is directly proportional to user satisfaction"*; 45% agreed, 31% remained neutral and 24% disagreed. Similarly, regarding the statement: *"The modularized system design results in operable software such that users encounter the difficulty levels* *gradually and progressively"* (Yunwen and Kishida, 2003), 52% agreed, 31% were neutral and 17% disagreed with the statement. The hypothesis *"H3: Operability plays a positive towards usability in OSS,"* as supported by the statistical analysis of our survey, has been accepted in our study.

About the attractiveness of software, 70% respondents of our survey believed that *"attractive to user"* software may not necessarily be a usable one, however 20% disagreed and the rest remained neutral. 67% agreed, 21% remained neutral and 12% disagreed that the more pleasant a software to use, more usable it would be. With the statement, *"Good user interface design is the result of properly applied usability techniques and practices"* (Chrusch, 2000), 79% agreed, 18% remained neutral and only 3% disagreed. Similarly, 94% believed that Usability is about *"total user experience,"* not only about attractive user interface (Markov, 2003); the percentages of the respondents who remained neutral and disagreed were 4% and 2% respectively. Our statistical analysis supports the hypothesis *"H4: Attractiveness is positively related with OSS usability,"* and is thus accepted in the study.

**Limitations of the Study & Threats to External Validity**

Empirical investigations of software engineering processes and products are done through surveys, experiments, metrics, case studies and field studies (Singer and Vinson, 2002). Wohlin et al. (2000) identify the ways in which the threats to external validity limit the researcher's ability to generalize the results of his/her experiment to industrial practice. In our study, we needed to support the external validity of our random sampling technique. Accordingly, we made a considerable effort to receive responses from many industry users; however, the total number of respondents was only 105 individuals. Although the proposed approach has some potential to threaten external validity, we followed appropriate research procedures by conducting and reporting tests to improve the reliability and validity of the study, and certain measures were also taken to ensure the external validity.

The increased popularity of empirical methodology in software engineering has also raised concerns on the ethical issues (Faden et al., 1986; Katz, 1972). We have followed the recommended ethical principles to ensure that the empirical investigation conducted and reported here would not violate any form of recommended experimental ethics.

## CONCLUSION

We have already conducted three studies to empirically investigate the significance of identified key factors on OSS usability from OSS developers, users and contributors' points of view. This research study is fourth of the series of our empirical investigations, in which we have analyzed the impact of the key usability factors (understandability, learnability, operability and attractiveness) on OSS usability based on industry users' perception. The key factors considered in the study (understandability, learnability, operability and attractiveness) have been taken from the standard ISO/IEC 9126-1 (2001). Empirical results of this study strongly support the hypotheses that understandability, learnability, operability and attractiveness have a positive impact on the usability of OSS projects. The study conducted and reported here shall enable OSS designers and developers to better understand the effectiveness of the relationships of the stated key factors and usability of their projects. Currently we are working on developing a maturity model to assess the usability of open source software projects. This empirical investigation provides us some justification to consider these key factors as measuring instruments.

**Appendix A. Key Usability Factors from OSS Industry Perspective (measuring instrument)**

**Understandability**: *"The capability of the software product to enable the user to understand whether the software is suitable, and how it can be used for particular tasks and conditions of use."* (ISO/IEC 9126-1, 2001)

1. To increase understandability in software functionality would have to be compromised.
2. Consistency in OSS design would increase understandability and hence usability.
3. Easy to understand software would encourage user's involvement.
4. Inconsistency in software is due to a lack of understanding user's expectations. (Chrusch, 2000)

**Learnability**: *"The capability of the software product to enable the user to learn its application."* (ISO/IEC 9126-1, 2001)

5. Learnability increases accessibility and hence usability.
6. In OSS environment, learnability may be compromised by developers for efficient products.
7. Learnability is a cognitive issue that needs users' mental analysis and is not practicable in OSS.
8. OSS developers must understand the limits of their target users to make a system learnable.

**Operability**: *"The capability of the software product to enable the user to operate and control it."* (ISO/IEC 9126-1, 2001)

9. More learnable software is more operable and hence usable.
10. Introduction of advance features of software to users in an incremental way would give user more control in using the software.
11. Operability is directly proportional to user satisfaction.
12. The modularized system design results in operable software such that users encounter the difficulty levels gradually and progressively. (Yunwen and Kishida, 2003).

**Attractiveness**: *"The capability of the software product to be attractive to the user."* (ISO/IEC 9126-1, 2001)

13. *"Attractive to user"* software may not necessarily be a usable one.
14. The more pleasant a software to use, more usable it would be.
15. Good user interface design is the result of properly applied usability techniques and practices. (Chrusch, 2000)
16. Usability is about *"total user experience,"* not only about attractive user interface. (Markov, 2003)

**Usability**: *"The capability of the software product to be understood learned, used and attractive to the user, when used under specified conditions."* (ISO/IEC 9126-1, 2001)

1. In OSS environment, adhering to standards and guidelines will take away OSS developer's freedom.

2. The adaptation of proven methods in OSS environment would ensure higher quality and address usability issues. (Hedberg et al., 2007)

3. In order to know end users' requirements and expectations, there is a need of more communication between the software developers and their target users, instead of relying on their instincts. (Koppelman and Van Dijk, 2006).

4. Usability increases development costs and lengthens development time. (Chrusch, 2000)